\documentclass[final,5p,tightenlines,times,twocolumn]{elsarticle}

\usepackage{amsmath}

\DeclareMathSymbol{\mathbbH}{\mathord}{AMSb}{"48}
\usepackage{amssymb}
\usepackage{bbold}
\usepackage{bm}
\usepackage{color,graphicx}
\usepackage{slashed}
\usepackage{comment}
\usepackage{hyperref}
\usepackage{subcaption}
\usepackage{makecell}
\usepackage[dvipsnames]{xcolor}
\usepackage{pifont}
\usepackage{caption}
\usepackage{float}
\usepackage{array}
    \newcolumntype{P}[1]{>{\centering\arraybackslash}p{#1}}
    \newcolumntype{M}[1]{>{\centering\arraybackslash}m{#1}}

\DeclareMathSymbol{\mathbbE}{\mathord}{AMSb}{"45}

\journal{Physics Letters B}

\begin{document}

\renewcommand\labelitemi{$\vcenter{\hbox{\tiny$\bullet$}}$}
\begin{frontmatter}
\title{Quantum entropy as a harbinger of factorizability}

\affiliation[anl]{%
   organization={High Energy Physics Division, Argonne National Laboratory},
   city={Lemont},
   postcode={IL 60439},
   country={USA}
}

\affiliation[nd]{%
   organization={Department of Physics and Astronomy, University of Notre Dame},
   city={Notre Dame},
   postcode={IN 46556},
   country={USA}
}

\author[anl,nd]{Henry Bloss}
\ead{hbloss@nd.edu}

\author[anl]{Brandon Kriesten}
\ead{bkriesten@anl.gov}

\author[anl]{T.J. Hobbs}
\ead{tim@anl.gov}

\begin{abstract}
Deeply inelastic scattering (DIS) is a powerful probe for investigating the QCD structure of hadronic matter and
testing the standard model (SM).
DIS can be described through QCD factorization theorems which separate contributions to the scattering
interaction arising from disparate scales --- {\it e.g.}, with nonperturbative matrix elements associated
with long distances and a perturbative hard scattering kernel applying to short-distance parton-level interactions.
The fundamental underpinnings of factorization may be recast in the quantum-theoretic terms of entanglement, (de)coherence, and system
localization in a fashion which sheds complementary light on the dynamics at work in DIS from QCD bound states.
In this Letter, we propose and quantitatively test such a quantum-information theoretic approach for dissecting factorization in
DIS and its domain of validity; we employ metrics associated with quantum entanglement such as a
differential quantum entropy and associated KL divergences in numerical tests.
We deploy these methods on an archetypal quark-spectator model of the proton, for which we monitor quantum decoherence in
DIS as underlying model parameters are varied.
On this basis, we demonstrate quantitatively how factorization-breaking effects may be imprinted on quantum
entropies in a kinematic regime where leading-twist factorization increasingly receives large corrections from finite-$Q^2$ effects;
our findings suggest potential applications of quantum simulation to QCD systems and their interactions.
\end{abstract}
\end{frontmatter}

%
%%%%%%%%%%%%%%%%
%%%%%%%%%%%%%%%%
%
\noindent ANL-195045

\section{Introduction}
\label{sec:intro}
Deeply inelastic scattering (DIS) is an essential source of information on hadron structure in high-energy phenomenology, playing
an important role in the precision of Standard Model (SM) baseline predictions at the LHC and other facilities. 
An array of DIS experiments contribute to the global data set for unfolding the parton distribution functions (PDFs)
of the proton~\cite{Hou:2019efy,Bailey:2020ooq,NNPDF:2021njg}, including both charged-lepton fixed-target and collider
measurements from BCDMS~\cite{BCDMS:1989qop}, JLab~\cite{Accardi:2023chb}, and HERA~\cite{H1:2015ubc}; meanwhile, neutrino DIS provides powerful
complementary information~\cite{Ruso:2022qes} and a range of future experiments have been proposed at DUNE~\cite{DUNE:2021tad}, the
EIC~\cite{AbdulKhalek:2021gbh}, and LHeC and FCC-eh~\cite{LHeC:2020van}.
These data can include measurements involving transverse momentum dependence relevant for the understanding of the multi-dimensional structure of hadrons~\cite{Soper:1979fq}
and complement probes sensitive to the nucleon's generalized parton distributions (GPDs)~\cite{Kriesten:2019jep,Kriesten:2021sqc}. 
The constraining power of DIS over PDFs and related quantities is closely tied to the predictivity of the QCD theory tested by the associated
data; this fact motivates the systematic examination of QCD over a broad kinematic range, particularly in regions where 
established theoretical frameworks become unstable.
In this Letter, we present a calculation grounded in QIS theory illustrating how basic concepts of entropy, localization, and
decoherence provide an intellectually rich foundation for such tests which might empower further investigations in quantum
simulation.

Specifically, we explore the validity and degeneration of factorization using quantum-theoretic measures as applied to a generic nonperturbative QCD model for DIS. 
In exploring the boundaries of factorization for this model scenario, we trace how long-distance correlations affecting hadronic structure functions are imprinted on the quantum entropy, offering insights into the dynamics of QCD processes and their interpretation; this work extends recent studies related to fragmentation functions~\cite{Benito-Calvino:2022kqa} and presents the full calculation previewed in recent $\nu$DIS proceedings~\cite{Bloss:2024bgr}.

\section{QCD factorization of DIS}
\label{sec:DIS}
At high energies, DIS is characterized at leading twist by the Compton-like scattering of a virtual boson from
partons which are effectively asymptotically free.
The short-distance part of the interaction can be expanded perturbatively in the strong coupling, $\alpha_{s}$, for which 
state-of-the-art calculations exist at N$^3$LO (see, {\it e.g.}, Ref.~\cite{Blumlein:2023aso} for a recent,
concise review).
Meanwhile, the long-distance remnant of the scattering process is encapsulated in nonperturbative matrix elements representing
quantum correlation functions --- in the case of spin-averaged DIS, these are the collinear, unpolarized PDFs. 
This separation of scales is formalized through QCD factorization theorems~\cite{Collins:2011zzd}, which hold to all orders
in $\alpha_s$; in unpolarized lepton-nucleon DIS, factorization may be applied to the DIS structure functions themselves,
leading to
\begin{eqnarray}
    \label{eq:fact}
F(x, Q^2) &=& \sum_{i,j}  \bigg\{ C_{i,j} \otimes f_j \bigg\}(x, Q^2)\, + \, \mathcal{O}\left({M^2 \over Q^2}\right)\ ,
\end{eqnarray}
where the short-distance Wilson coefficient functions, $C_{i,j}$, can be calculated perturbatively and
$f_j$ contain the nonperturbative PDFs of a nucleon of mass, $M$; here, the indices $\{i,j\}$ are parton-flavor labels, and the variables $x$ and $Q^{2}$ specify the external kinematics via the Bjorken invariant (identified with the longitudinal momentum
fraction carried by the struck parton) and DIS resolution scale, respectively.
At the kinematical boundaries of the validity of Eq.~(\ref{eq:fact}), the $\sim\!\! 1/Q^2$ power-suppressed contributions
which correct the first term begin to spoil the separation of short- {\it vs.}~long-distance scales which underpin factorizability.

A desire to systematically explore the kinematical dependence of such $\sim\!\! 1/Q^2$ effects which cloud
the theoretical interpretation of DIS measurements therefore motivates empirical tests of factorization.
In principle, factorization-breaking effects arise from various kinematical and dynamical mechanisms, which might be suppressed
by a range of characteristic internal scales, for instance, $\sim\! M^2/Q^2$ or $\sim\! k^2_T/Q^2$, with the latter related to intrinsic quark transverse
momentum.
Moreover, radiative effects associated with soft-gluon emission which might ordinarily be resummed into quantum correlation functions might
instead interfere with the partonic hard scatter for $Q^2\! \sim\! \langle k^2_T \rangle$; these correspond to long-range correlations which are
explicitly non-factorizable.

\begin{figure}
    \centering
    \includegraphics[width=\linewidth]{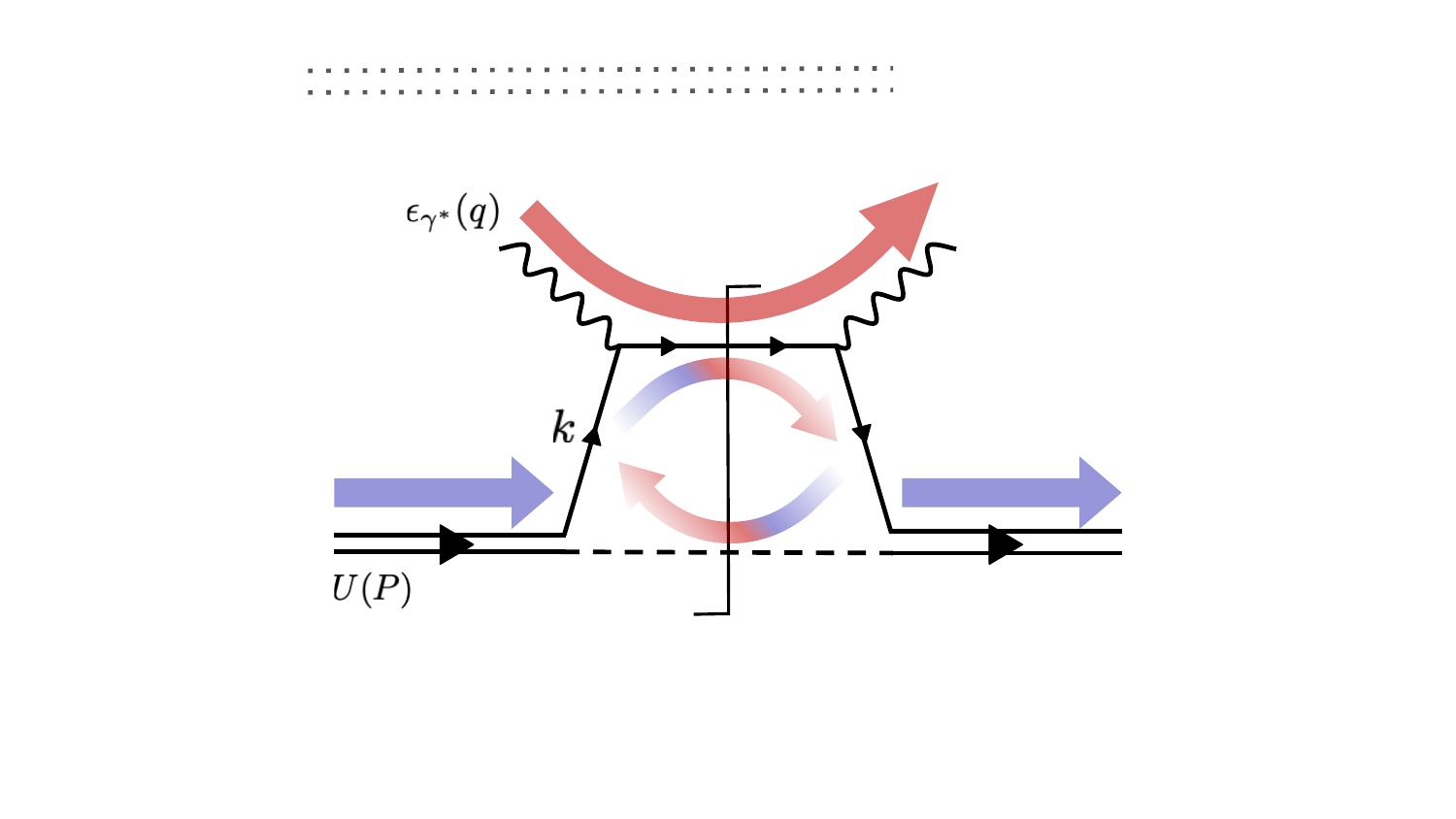}
    \caption{The DIS handbag diagram in a nonperturbative quark-diquark model, with colored arrows representing separate flows of quantum-theoretic information; the blue arrows highlight the information contained in the long-distance hadronic process while the red arrow denotes the information associated with the short-distance partonic scattering.}
    \label{fig-1}
\end{figure}

In light of this incomplete understanding of the possible mechanisms which might spoil factorization and their corresponding
dependence on $x$ and $Q^{2}$, theoretical models for DIS with and without the explicit assumption
of factorization are especially valuable.
A class of such models possessing the minimal requirements of exact
calculability and nonperturbative behavior reflected by internal scales furnished by constituent quark masses and $k_T$ are quark-spectator pictures of the proton~\cite{Kriesten:2019jep,Kriesten:2021sqc,Hobbs:2013bia,Hobbs:2014lea,Hobbs:2016xfz,Hobbs:2017fom,Moffat:2017sha}.
Without loss of generality, we adopt the assumptions of Ref.~\cite{Moffat:2017sha}, which employed a spectator-diquark model with a nucleon-quark-diquark interaction of the form, $\mathcal{L}_{\text{int}} = -\lambda \bar{\Psi}_N \psi_q \phi_D + \textit{h.c.}$; this effective interaction gives
rise to the lower vertices in the DIS handbag diagram depicted in Fig.~\ref{fig-1}. 
The advantage of this approach is that it is then possible to compute DIS structure functions consistently and exactly from the appropriate diagrams of the type shown in Fig.~\ref{fig-1}, in addition to parallel calculations which explicitly assume factorization along the lines of Eq.~(\ref{eq:fact}).

%
%%%%%%%%%%%%%%%%
%%%%%%%%%%%%%%%%
\section{Quantum information-theoretic metrics for factorization breaking}
\label{sec:entropy}

%%%
Recent years have seen a renaissance in reframing fundamental processes in particle physics, including high-energy
interactions with QCD bound states, in a quantum-statistical language involving measures formally derived
from (de)coherence, entanglement entropy, and related concepts~\cite{Kharzeev:2017qzs,Berges:2017zws,Shuryak:2017phz,Aidala}.
%
%%%
In an information-theoretic sense, factorization relies on the ability to disentangle distinct quantum subsystems and their associated
dynamics. In Fig.~\ref{fig-1}, this concept is depicted schematically alongside the DIS handbag, over which the blue arrows represent the
flow of information in the long-distance binding of the interacting parton into the asymptotic hadronic state, while the red arrow
represents the information carried in the short-distance coupling of the external boson at parton-level.

Eq.~(\ref{eq:fact}) suggests that factorizability naturally rests on a foundation of quantum decoherence~\cite{Aidala},
which permits the separation of long-distance physics encoded in the PDFs ($f_j$) and the short-distance interactions of the
partonic-level cross section ($C_{i,j}$), up to the controllable $\sim\!1/Q^2$ corrections discussed in Sec.~\ref{sec:DIS}. 
In this picture, any residual correlations among the subprocess in the boson-nucleon DIS interaction generate some entanglement 
which spoils the decoherence among short- and long-distance contributions manifest in the factorized expression of Eq.~(\ref{eq:fact}).
The entanglement seeded by these decoherence-spoiling correlations in turn leave a signature on quantum information-theoretic
quantities such as the differential entropy, and serve as a proxy for the breakdown of QCD factorization.
In models like Ref.~\cite{Moffat:2017sha}, these effects enter as, {\it e.g.}, $\sim\! k^2_T / Q^2$ deviations between the factorized and exact expressions and can be computed systematically.
Similar formulations of entanglement entropy have been calculated~\cite{Hagiwara:2018sha} on the fully unintegrated, multi-dimensional wave function of the proton --- the Wigner distribution --- from the Wehrl entropy~\cite{WEHRL1979353}.

%%%
Using a quark-diquark model as in Ref.~\cite{Moffat:2017sha}, we compute $k_T$-dependent structure functions defined as
\begin{equation}
\label{eq:norm}
    F_1(x,Q^2) = {1 \over (2\pi)^2}\int d^{2}\mathbf{k}_{T}\ f_1(x, k_T, Q^2)\ ,
\end{equation}
with $k_T = |\mathbf{k}_{T}|$, such that the unintegrated, $k_{T}$-dependent structure function, $f_1(x, k_T, Q^2)$, permits an interpretation as a statistical distribution.
In the discussion below, we normalize $f_1(x, k_T, Q^2)$ by hand according to Eq.~(\ref{eq:norm}).
On this basis, we define a {\it differential entropy} associated with the $k_T$ distribution~\cite{Hagiwara:2018sha} over the model structure function, $F_1$; namely,
\begin{equation}
\label{eq:entropy_int}
    H_{F_1}(x,Q^2) = - {1 \over (2\pi)^2} \int d^{2}\mathbf{k}_{T}\ f_1(x, k_T, Q^2)\, \ln \left[f_1(x, k_T, Q^2)\right]\ ;
\end{equation}
to gain insights into this quantity, we explore its $k_T$-dependent integrand --- specifically,
\begin{equation}
\label{eq:entropy}
    \mathcal{H}_{F_1}(x, k_T, Q^2)\ \equiv\ -k_T f_1(x, k_T, Q^2)\, \ln \left[ f_1(x, k_T, Q^2) \right]\ .
\end{equation}
While care is needed in the interpretation of entropies defined over continuous distributions as above, the differential entropy is mathematically connected to notions of localization as reflected in the shape of the underlying densities --- for instance, more localized systems generally produce sharply peaked distributions with correspondingly strongly negative differential entropies.
As Eq.~(\ref{eq:entropy}) emerges from an inherently quantum mechanical object, these indications of localization may be connected in a model context to the decoherence of the associated wave function.

%%%
To track their differences, we also evaluate {\it relative entropies} of the factorized and exact calculations to assess the shift in probability mass from one scenario to the other. 
The appropriate quantity is the KL divergence between the factorized and exact entropy distributions, and can be defined as
\begin{equation}
\label{eq:kl_div}
    \mathcal{D}_{KL}(F_1^{\text{exact}} \| F_1^{\text{fact}})\ \equiv\ k_{T} f_{1}^{\text{exact}}(x, k_{T}, Q^{2}) \ln \left[\frac{f_{1}^{\text{exact}}(x, k_{T}, Q^{2})}{f_{1}^{\text{fact}}(x, k_{T}, Q^{2})} \right]\ ;
\end{equation}
we note that similar KL divergences have attracted in interest in past investigations of partonic densities and fragmentation
functions~\cite{Benito-Calvino:2022kqa}.
We point out that this quantity is not symmetric and that the probability mass to go from exact to factorized (forward) is not the same, by definition, as the probability mass to go from factorized to exact (backward). 
The KL divergence in the forward case gives us a quantitative measure of the information (in relevant ``bits'' units) needed to describe the exact calculation when starting from the factorized scenario. 
In other words, the relative entropy quantifies the information cost associated with the assumption of factorization in the theoretical description of DIS. 
This notion is connected to the fact that the KL divergence may be described through an information-theoretic cross-entropy as
\begin{eqnarray}
    \mathcal{D}_{KL} (F_1^{\text{exact}} \| F_1^{\text{fact}}) = \mathbbH(F_1^{\text{exact}} , F_1^{\text{fact}}) - \mathbbH(F_1^{\text{exact}} )\ ,
\end{eqnarray}
such that large values of the KL divergence are associated with high information cost or cross-entropy, as compared with the uncertainty underlying the initial distribution; this implies that the factorized calculation fails to capture entangling correlations present in the exact calculation in the implicated regions of $k_{T}$.
Having presented the concise formal basis for our calculation, we illustrate the application to the quark-diquark model in Sec.~\ref{sec:results} below.

%%%%%%%%%%%%%%%%
%%%%%%%%%%%%%%%%
\section{Results}
\label{sec:results}

%%%
In our numerical calculations, we target the boundary region of factorizability: moderately low-$Q^{2}$, $Q^2 = 4$ GeV$^2$, and large Bjorken-$x$ values near $x_{Bj} =0.6$. 
In addition, we select several typical values for the mass parameters, taking $m_{q} = 0.3, 0.5$ GeV for the constituent quark masses and allowing the scalar spectator mass to vary over the range $m_{s} = 0.65\! -\! 0.75$ GeV, similarly to the assumptions in Ref.~\cite{Moffat:2017sha}.
These choices are motivated by their sensitivity to the emergence of factorization-breaking effects and are phenomenologically relevant to the kinematics associated with soft DIS as discussed in the neutrino-focused proceedings of Ref.~\cite{Bloss:2024bgr}. 
Moreover, this kinematic region lies near the initial parameterization scale of the PDFs where there are significant parameterization uncertainties \cite{Courtoy:2022ocu,Kotz:2023pbu}.
It may therefore be possible to explore connections to analogous uncertainties in phenomenological fits through such information theoretic measures --- we leave such questions to future study.

%%%
With these choices, we compute the differential entropies of the continuous DIS structure functions as summarized in Sec.~\ref{sec:entropy} above, evaluating both the $k_T$-(un)integrated objects of Eqs.~(\ref{eq:entropy_int}) and (\ref{eq:entropy}). 
These quantities are evaluated for both the exact and factorized scenarios of the model, such that we quantify the impact of factorization-breaking effects on the entropy associated with the $k_T$ distributions.
We compute each scenario assuming four combinations of constituent mass parameters representing modest variations in ratios among the internal model scales, $m^2_q/Q^2$, {\it etc}.

\begin{figure}[!ht]
    \centering
    \includegraphics[width=\linewidth]{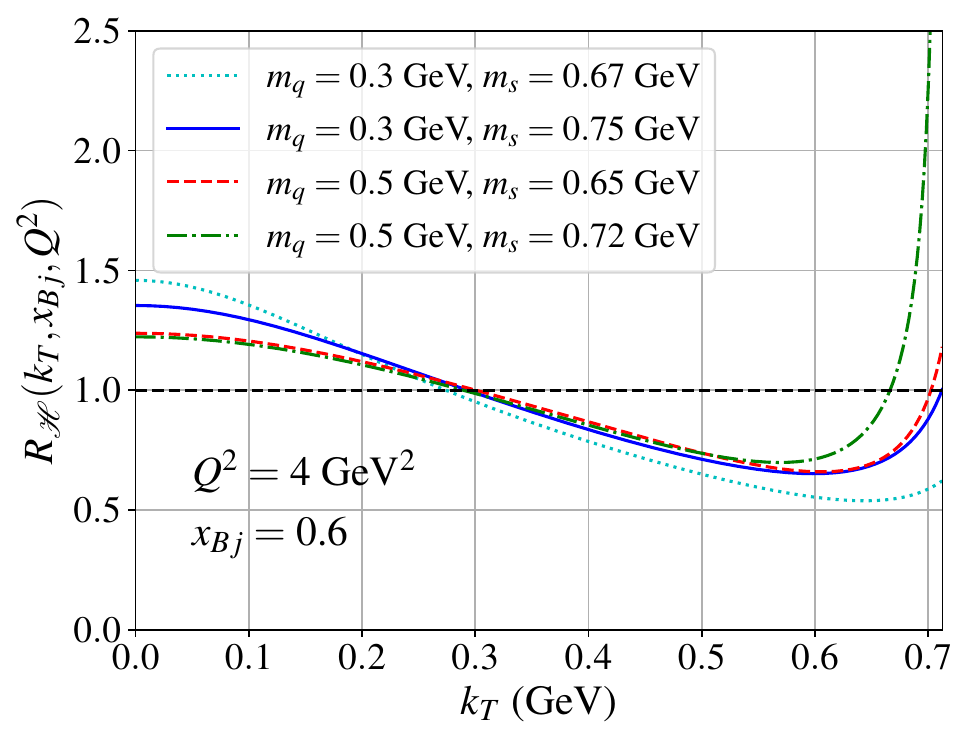}
    \includegraphics[width=\linewidth]{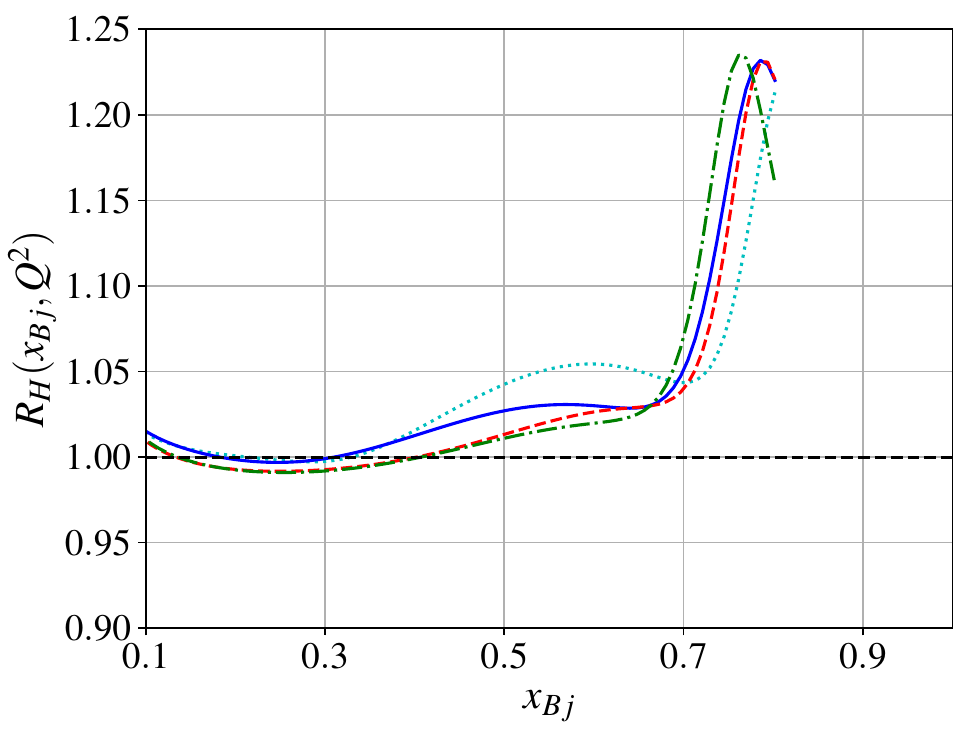}
    \caption{(Top) Ratio of the $k_{T}$-dependent differential entropies of factorized to exact for several different nonperturbative model scenarios associated with varied constituent mass parameters. (Bottom) Ratio of $x_{Bj}$ dependent fully integrated entropies of factorized and exact for the same constituent mass parameters. Both calculations are performed at a scale of $Q^{2} = 4$ GeV$^{2}$.}
    \label{fig-2}
\end{figure}
%

%%%
In Fig.~\ref{fig-2} (top), we plot the ratio of the $k_T$-dependent entropy as defined in Eq.~(\ref{eq:entropy}) for the exact calculation relative to the factorized case.
The ratio highlights the effects of the factorization-breaking mechanisms present in the exact calculation --- which in this model scenario can be as large as $50\%$ in some $k_T$ regions. 
Similarly, in the bottom panel, we plot the corresponding ratio of the $k_T$-integrated differential entropy. 
The differential entropy is allowed to be negative due to the nature of the continuous distributions on which they are evaluated, especially in regions of greater localization of the system in configuration space. 
As such, the ratios shown in Fig.~\ref{fig-2} represents relative shape variations in the {\it magnitude} of the entropies.

%%%
%
Inspecting Fig.~\ref{fig-2} (top), we see that the factorization-breaking effects present in the exact calculation drive a relative increase in the $k_T$-dependent contribution to the entropy (making it more strongly negative), while suppressing it at larger values of $k_T$. 
In Fig.~\ref{fig-2} (bottom), we find that these $k_T$ dependent contributions are reflected in the corresponding integrated differential entropy of Eq.~(\ref{eq:entropy_int}), such that the relative entropies of the exact and factorized calculations diverge at high $x_{Bj}$; this behavior signifies the growing role of factorization-breaking dynamics for modest but small fixed $Q^2$; we note that these effects become still larger with diminishing values of $Q^2$.
The relative increase in these entropies indicates residual coherence which degrades the purely leading-twist factorization of Eq.~(\ref{eq:fact}).

%%%
Fluctuating the size of factorization-breaking effects by varying constituent masses relative to the hard scale, $Q^2$, has a mild effect in the intermediate ranges of $k_T$ and $x$ shown here, but these effects grow steadily deeper into the boundaries of the factorizable regime. 
For example, in Fig.~\ref{fig-2} (top) this effect can be seen clearly, especially as $k_{T}$ approaches larger values comparable to $Q$. 
Notably, it is in this comparatively large-$k_T$ region that variations in the internal model scales ({\it i.e.}, the masses) most affect the differential entropy.
Similarly in Fig.~\ref{fig-2} (bottom), a kinematic threshold is apparent in the $k_T$-integrated differential entropy, whereat the factorization-breaking effects strongly escalate.

\begin{figure}
    \centering
    \includegraphics[width=\linewidth]{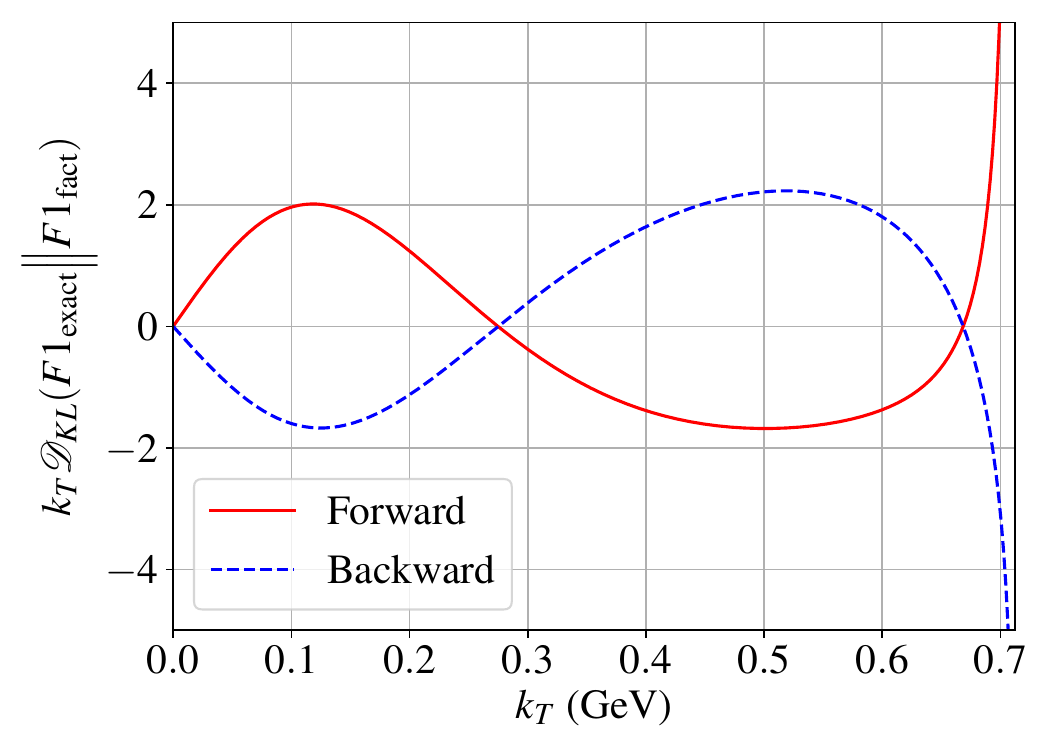}
    \caption{The $k_{T}$-dependent KL-divergence calculated at $Q^2 = 4$ GeV$^2$ and $x_{Bj} = 0.6$ in two schemes; forward in which the KL divergence is evaluated as exact versus factorized, and backward where the the KL divergence is evaluated as factorized versus exact.}
    \label{fig-3}
\end{figure}

%%%
Finally, in Fig.~\ref{fig-3} we calculate the $k_{T}$-dependent KL-divergence defined in~Eq.~\ref{eq:kl_div} at the same scale, $Q^{2} = 4$ GeV$^{2}$. 
We perform the calculation of this quantity in two standard implementations, labeled ``forward'' and ``backward,'' following common notation in information theory.
Both cases characterize how the probability mass shifts when one distribution is used to approximate the other. 
Consequently, the regions where there are large deviations from zero represent substantial information loss, reflecting, in an equivalent bit-unit, the removal of dynamical content contained in the exact calculation upon going to the factorized approximation. 
This information loss can be interpreted as the ``information cost'' associated with the onset of factorization and therefore, the loss of decoherence.

%
%
%%%%%%%%%%%%%%%%
%%%%%%%%%%%%%%%%
\section{Conclusions}

%%%
In this Letter, we have applied quantum information theoretic concepts and measures, especially in the form of a differential quantum entropy, to demonstrate the connection between the loss of decoherence and the breakdown of QCD factorization in DIS. 
Our results indicate that there is a description of factorization of interactions with QCD bound states through a quantum mechanical localization of the system.
Further, that these entropies are particularly sensitive to the mechanism driving the factorization --- suggesting a possible probe for future studies in experimental programs near these thresholds.

%%%
Further refinement of this formalism suggests an opportunity to directly extract information from quantum simulations of QCD bound states; for example, entropies which might be directly calculable in quantum simulations~\cite{Khor:2023xar} may provide independent information on the formation of structure or interpretation of dynamics which break fundamental QCD relations like factorization theorems.

%%%
The connection between information theory and AI/ML methods such as entropy metrics and uncertainty quantification~\cite{Kriesten:2024ist} suggest that information theoretic quantities as relevant to generative AI might serve as a plausible testbed for studying factorization and related fundamental aspects of QCD.
Furthermore, the embedding of DIS factorization theorems into a broader framework of collinear QCD interactions allows for the inclusion of not just inclusive but also exclusive structure functions~\cite{Almaeen:2022imx, Kriesten:2023uoi, Kriesten:2024are, Almaeen:2024guo} --- effectively looping in GPDs and their description of the spatial structure and charge localization.

%%%%%%%%%%%%%%%%
%%%%%%%%%%%%%%%%
%
%

\section{Acknowledgments}
The work of Henry Bloss was supported by the U.S.~Department of Energy, Office of Science, Office of Workforce Development for Teachers and Scientists (WDTS) under the Science Undergraduate Laboratory Internship (SULI) program.
The work of Brandon Kriesten and T.J. Hobbs at Argonne National Laboratory was supported by the U.S.~Department of Energy under contract DE-AC02-06CH11357.
%
%

%%%%%%%%%%%%%%%%
%%%%%%%%%%%%%%%%
%
%

\bibliographystyle{elsarticle-num} 
\bibliography{qfact}

\end{document}